\documentclass[12pt,amsfonts,amsmath,amssymb,color]{article}
\usepackage{authblk}
\def\baselinestretch{1.0}
\usepackage{amsmath,amssymb,amsfonts,latexsym,float,graphics,epsfig}
\usepackage{subfig}
\usepackage{verbatim}
\usepackage{relsize}
\usepackage{hyperref}
\usepackage{graphicx}
\usepackage{bm}
\usepackage{epstopdf}
\usepackage{slashed}
\usepackage{color}
\usepackage{tikz-cd}
\usepackage[utf8]{inputenc}
\usepackage[T1]{fontenc}

\begin{document}

\renewcommand\theequation{\arabic{section}.\arabic{equation}}
\catcode`@=11 \@addtoreset{equation}{section}

\newtheorem{axiom}{Definition}[section]
\newtheorem{theorem}{Theorem}[section]
\newtheorem{axiom2}{Example}[section]
\newtheorem{lem}{Lemma}[section]
\newtheorem{prop}{Proposition}[section]
\newtheorem{corollary}{Corollary}[section]

\let\endtitlepage\relax

\begin{titlepage}
\begin{center}
\renewcommand{\baselinestretch}{1.5}

{\Large \bf{Generalized virial theorem}}\\
 {\Large \bf{for contact Hamiltonian systems}}

\vspace{9mm}
\renewcommand{\baselinestretch}{1}  

\centerline{\large{\bf Aritra Ghosh\footnote{ag34@iitbbs.ac.in, ~aritraghosh500@gmail.com}}}

\vspace{5mm}
\normalsize
\text{School of Basic Sciences,}\\
\text{Indian Institute of Technology Bhubaneswar,}\\
\text{Argul, Jatni, Khurda, Odisha 752050, India}\\
\vspace{5mm}

\begin{abstract}
We formulate and study a generalized virial theorem for contact Hamiltonian systems. Such systems describe mechanical systems in the presence of simple dissipative forces such as Rayleigh friction, or the vertical motion of a particle falling through a fluid (quadratic drag) under the action of constant gravity. We find a generalized virial theorem for contact Hamiltonian systems which is distinct from that obtained earlier for the symplectic case. The `contact' generalized virial theorem is shown to reduce to the earlier result on symplectic manifolds as a special case. Various examples of dissipative mechanical systems are discussed. We also formulate a generalized virial theorem in the contact Lagrangian framework. 
\end{abstract}
\end{center}
\end{titlepage}

\section{Introduction}
The virial theorem, introduced by Rudolf Clausius \cite{vt1} has found applications in astrophysics, cosmology, molecular physics, quantum mechanics and statistical mechanics (see for instance \cite{vt2,vt3,vt4,vt5,hv}). In the context of classical mechanics, the virial theorem relates the time-averaged kinetic energy $\langle K \rangle$ with the time-average of $\mathbf{r} \cdot \mathbf{F}$, where $\mathbf{r}$ is the position vector while $\mathbf{F}$ is the force vector. In his original paper \cite{vt1}, Clausius introduced the `virial' for a one-particle system, defined as $G = m \mathbf{r}\cdot \mathbf{v}$ and showed that if the motion is periodic in time $T$ or at least if $G$ remains bounded in its time evolution, then 
\begin{equation}\label{1st}
\langle K \rangle = -\frac{1}{2} \langle \mathbf{r} \cdot \mathbf{F} \rangle,
\end{equation}
where angled brackets $\langle \cdot \rangle$ denote `time-averaging'. In particular, if the force is conservative, one can write $\mathbf{F} = -\nabla V$ and therefore
\begin{equation}
\langle K \rangle = \frac{1}{2} \langle \mathbf{r} \cdot \nabla V \rangle.
\end{equation}
This result can be straightforwardly extended for quantum mechanical systems \cite{vt5,hv} and the tensor virial equations \cite{vt2} which indicate that the kinetic and potential energies must be in `balance' in each separate direction. A generalized version of the virial theorem, called the `hypervirial' theorem was studied in \cite{hv} and certain hypervirial relations have been recently obtained in \cite{hvr}. The formulation of the hypervirial theorem or simply the generalized virial theorem was studied in \cite{virial1,virial2,virial3,virial4} using methods of symplectic geometry, wherein various mechanical examples including position-dependent mass systems were discussed. The purpose of this note is to study the generalized virial theorem for contact Hamiltonian systems which describe some simple dissipative systems in mechanics. 

\vspace{2mm}

Contact geometry is like an odd-dimensional cousin of symplectic geometry \cite{goldsteincm,arnoldcm} with several common features, such as the existence of Hamiltonian dynamics \cite{arnoldcm,Geiges,Arnold} as well as invariants analogous to Gromov-Witten invariants \cite{El}. However, unlike their symplectic counterparts, Hamiltonian dynamics on a contact manifold is neither volume preserving, nor does it conserve the corresponding Hamiltonian function along the evolution. Consequently, contact Hamiltonian dynamics has found applications in describing dissipative mechanical systems \cite{CM1,CM2,CM3,CM4,CM5,CM6,CM7,CM8}, as well as thermostat problems where the system interacts with an environment \cite{thermo}. It has also found applications in reversible \cite{hermann,RT1,RT2,RT3,RT4,RT5,RT6,RT7,RT8,Peter79} as well as irreversible thermodynamics \cite{generic1,generic2,generic4}. Given these features, it appears interesting to explore the generalized virial theorem for a contact Hamiltonian system which may describe a dissipative system such as that experiencing linear or quadratic drag forces where the kinetic and potential energies are not conserved. Furthermore, we also obtain a generalized virial theorem in the contact Lagrangian framework where the dynamics coincides with that described by the Herglotz variation problem, originally presented more than 90 years back \cite{Herglotz} (see also \cite{CM4} for a nice review). Various examples are discussed throughout the paper. 

\vspace{2mm}

With the above background, we present the organization of this paper as follows. In the next section [Sec. (\ref{IntroSec})], we will introduce the reader to some basic contact geometry and will fix our notation. The role of antisymmetric brackets is emphasized upon. Following this, in Sec. (\ref{VirSec}), we present a generalized virial theorem for contact Hamiltonian systems and show that the earlier result \cite{virial1,virial4} arises only as a special case. Some examples are discussed. Then, in Sec. (\ref{LagSec}), we will derive another generalized virial theorem from the framework of contact Lagrangian dynamics. Finally, we conclude the paper in Sec. (\ref{DisSec}). 

\section{Contact geometry}\label{IntroSec}
In this section, we provide a minimal background of Hamiltonian dynamics on contact manifolds. We begin by describing symplectic manifolds first. 

\subsection{Symplectic manifolds}
Let us recall that a symplectic manifold is a pair $(\mathcal{M}_s,\omega)$ where $\mathcal{M}_s$ is a smooth manifold of real dimension $2n$ and $\omega$ is a closed and non-degenerate two-form, i.e. it satisfies
\begin{equation}
  d\omega = 0, \quad \quad \omega^n \neq 0,
\end{equation} where $\omega^n = \omega^{\wedge n} = \omega \wedge \omega \wedge \cdots \wedge \omega$ ($n$ times) is the volume form on $\mathcal{M}_s$. Further, a symplectic manifold $(\mathcal{M}_s,\omega)$ shall be called exact if $\omega = d\theta$, where $\theta$ is a one-form called the symplectic potential. 

\vspace{2mm}

Now, for any function $H \in C^\infty(\mathcal{M}_s,\mathbb{R})$, one has the vector-bundle map between $T\mathcal{M}_s$ and $T^*\mathcal{M}_s$:
\begin{equation}\label{symplecticHamiltonian}
  \iota_{X_H} \omega = dH,
\end{equation} where $X_H$ is a vector field determined by the above condition. Note that since $\omega$ is non-degenerate, this map is an isomorphism. The vector field $X_H$ is known as the Hamiltonian vector field corresponding to the Hamiltonian function $H$. These vector fields are volume preserving which can be seen by constructing the Lie derivative of $\omega$ with respect to $X_H$:
\begin{equation}
  \pounds_{X_H} \omega = d (\iota_{X_H} \omega) + \iota_{X_H} (d\omega) = 0.
\end{equation} This implies $ \pounds_{X_H} \omega^n = 0$ thereby leading to the familiar Liouville's theorem used in statistical mechanics. There is a theorem due to Darboux stating that on a symplectic manifold, near any point, there are local coordinates $(q^i,p_i)$ such that the two-form $\omega$ reads
\begin{equation}\label{symplecticDarboux}
  \omega = dq^i \wedge dp_i.
\end{equation}
It can be easily checked that the local expression of $X_H$ consistent with Eqs. (\ref{symplecticHamiltonian}) and (\ref{symplecticDarboux}) is
\begin{equation}
  X_H = \frac{\partial H}{\partial p_i} \frac{\partial }{\partial q^i} - \frac{\partial H}{\partial q^i} \frac{\partial}{\partial p_i}   .
\end{equation}
Then, clearly the integral curves of $X_H$ satisfy the Hamilton's equations of motion. The function $H$ is conserved under the flow generated by $X_H$, i.e. $X_H (H) = 0$. For simple mechanical systems where $H$ is the energy, this leads to conservation of energy. 

\subsection{Contact manifolds}
A contact manifold is a pair $(\mathcal{M}_c,\eta)$ where $\mathcal{M}_c$ is a smooth manifold of real dimension $2n+1$ and $\eta$ is a one-form satisfying
\begin{equation}\label{nonintegrability}
  \eta \wedge (d\eta)^n \neq 0,
\end{equation} which is known as the condition of maximal non-integrability. Here, $(d\eta)^n = (d\eta)^{\wedge n}$. Note that $\eta \wedge (d\eta)^n$ is the considered volume form on $\mathcal{M}_c$. In the context of Frobenius integrability, Eq. (\ref{nonintegrability}) means that the hyperplane distribution defined as ${\rm ker}(\eta)$ is maximally non-integrable in the sense that the resulting hyperplanes are overly twisted. We refer the reader to \cite{arnoldcm,Geiges} for more details.

\vspace{2mm}

On any contact manifold $(\mathcal{M}_c,\eta)$, there exists a global vector field $\xi$ known as the Reeb vector field defined uniquely through the relations\footnote{Contact manifolds may be associated with an additional metric structure, in the sense of Sasaki \cite{sasaki}. Then the first amongst Eq. (\ref{xidef}) indicates that $\xi$ appears as a dual vector to $\eta$ as  $g_{ij} \xi^j = \eta_i$.}
\begin{equation}\label{xidef}
  \eta(\xi) = 1, \quad \quad \iota_\xi d\eta = 0.
\end{equation} There is an analogous Darboux's theorem for contact manifolds which says that near any point, it is possible to define local (Darboux) coordinates $(s,q^i,p_i)$ such that
\begin{equation}
  \eta = ds - p_i dq^i,
\end{equation} and thus $\xi = \partial/\partial s$ in local coordinates. Let us note that since locally $d\eta = dq^i \wedge dp_i$, one may think of a contact manifold to be locally of the form $\mathcal{M}_c = \mathcal{M}_s \times \mathbb{R}$, where $\mathcal{M}_s$ is a (exact) symplectic submanifold (codimension one) of $\mathcal{M}_c$ such that $\theta = \Phi^* \eta$ is the symplectic potential on $\mathcal{M}_s$, for the inclusion map $\Phi:\mathcal{M}_s \mapsto \mathcal{M}_c$.

\subsection{Contact Hamiltonian dynamics}\label{chd}
With the basic understanding of contact manifolds as introduced above, let us describe Hamiltonian dynamics analogous to the symplectic case. For a function $h \in C^\infty(\mathcal{M}_c,\mathbb{R})$, there is an associated vector field $X_h$ defined by the combined conditions
\begin{equation}\label{contactvecdef}
  \eta(X_h) = -h, \quad \quad \iota_{X_h} d\eta = dh - \xi(h) \eta.
\end{equation} The vector field $X_h$ is known as the contact Hamiltonian vector field associated with the function $h$ and in local (Darboux) coordinates, it takes the following form:
\begin{equation}
  X_h = \bigg(p_i \frac{\partial h}{\partial p_i} - h \bigg) \frac{\partial }{\partial s} + \bigg(\frac{\partial h}{\partial p_i}\bigg)\frac{\partial }{\partial q^i} - \bigg( \frac{\partial h}{\partial q^i} + p_i \frac{\partial h}{\partial s} \bigg) \frac{\partial}{\partial p_i}.
\end{equation}
Clearly, this vector field does not conserve $h$ along its flow, i.e.
\begin{equation}
  X_h (h) = - h \frac{\partial h}{\partial s} \neq 0,
\end{equation}
which may also be seen without referring to the local (Darboux) coordinate expressions just by contracting the second amongst Eq. (\ref{contactvecdef}) with $X_h$ and then using the first one. Furthermore, the flow is not volume preserving because 
\begin{equation}
\pounds_{X_h} (\eta \wedge (d\eta)^n) = {\rm div}.X_h  (\eta \wedge (d\eta)^n),
\end{equation} where the divergence of $X_h$ is found to be 
\begin{equation}
  {\rm div}.X_h = -(n+1) \frac{\partial h}{\partial s} = -(n+1) \xi(h).
\end{equation} Thus, some of the nice properties of (symplectic) Hamiltonian vector fields are not present in their contact-geometric counterparts. However, it is very easy to see that if $h$ becomes independent of $s$, then the contact vector field acquires all the properties of a symplectic vector field because $\xi(h) = 0$. 

\subsection{Time-dependent contact dynamics}\label{TDsec}
Let us briefly describe the situation where $h$ may carry an explicit time dependence (see also \cite{CM2,CM7,CM8}). Such cases lead to interesting dynamics and have been discussed in Secs. (\ref{Forsec}) and (\ref{Bsec}). Let us consider an extended phase space $M' = \mathcal{M}_c \times \mathbb{R}$ where $\mathcal{M}_c$ is a contact manifold on which $(s,q^i,p_i)$ are local (Darboux) coordinates and $t \in \mathbb{R}$ (global coordinate) such that $h \in C^\infty(M', \mathbb{R})$. Consider the one-form $\alpha = h dt + ds - p_i dq^i$ on $M'$, being defined as $\eta = \Psi^* \alpha$ where $\Psi: \mathcal{M}_c  \mapsto M'$. Then, we define the time-dependent evolution vector field $\Xi_h$ via the following intrinsic conditions (see also \cite{CM2}):
\begin{equation}
\alpha(\Xi_h) = 0, \quad \quad \iota_{\Xi_h} d\alpha = \zeta(h) dt - \xi(h) \alpha,
\end{equation}
 where $\zeta = \partial / \partial t$ is a global vector field on $M'$. Subsequently, in coordinates $(t,s,q^i,p_i)$ on $M' = \mathcal{M}_c \times \mathbb{R}$, we have
 \begin{equation}
 \Xi_h =  \frac{\partial}{\partial t} + X_h.
 \end{equation}
 The resulting equations of motion are the same as those obtained in Sec. (\ref{chd}) along with $\dot{t} = 1$.

\subsection{Lagrange (Jacobi) brackets}
Contact manifolds fall into the broader class of Jacobi manifolds \cite{deLeon} (see also \cite{jacobi1,jacobi2}) which are equipped with a local Lie-bracket structure $\{\cdot,\cdot\}:C^\infty(\mathcal{M}_c,\mathbb{R}) \times C^\infty(\mathcal{M}_c,\mathbb{R}) \rightarrow C^\infty(\mathcal{M}_c,\mathbb{R})$. In general, a Jacobi structure is the triple $(M,\Lambda,E)$ where $M$ is a smooth manifold (not necessarily contact) while $\Lambda$ is a bi-vector field and $E$ is a vector field satisfying the conditions:
\begin{equation}\label{integrabilitycond}
[\Lambda, \Lambda] = 2 E \wedge \Lambda, \quad \quad [\Lambda, E] = 0,
\end{equation} where $[\cdot, \cdot]$ is the Schouten-Nijenhuis bracket \cite{SN}. Such a manifold $M$ admits a local Lie bracket $\{\cdot,\cdot\}:C^\infty(M,\mathbb{R}) \times C^\infty(M,\mathbb{R}) \rightarrow C^\infty(M,\mathbb{R})$ given by
\begin{equation}
\{ f , g \} = \Lambda (df,dg) + f E (g) - E(f) g,
\end{equation} for $f, g \in C^\infty(M, \mathbb{R})$. The bracket is called a Jacobi bracket which is clearly antisymmetric and also satisfies the Jacobi identity (due to Eq. (\ref{integrabilitycond})).

\vspace{2mm}

On a contact manifold, i.e. with $M = (\mathcal{M}_c,\eta)$, we identify $E = \xi$ (Reeb vector field) and $\Lambda (df,dg) = d\eta(X_f,X_g)$ which gives in Darboux coordinates, the following expression: 
\begin{eqnarray}
  \{f,g\} &=& f \frac{\partial g}{\partial s} - \frac{\partial f}{\partial s} g + p_i \bigg( \frac{\partial f}{\partial s} \frac{\partial g}{\partial p_i} - \frac{\partial f}{\partial p_i} \frac{\partial g}{\partial s}\bigg) + \frac{\partial f}{\partial q^i} \frac{\partial g}{\partial p_i} - \frac{\partial f}{\partial p_i} \frac{\partial g}{\partial q^i},
\end{eqnarray} for any $f , g \in C^\infty(\mathcal{M}_c,\mathbb{R})$. This Jacobi bracket on a contact manifold has been called the Lagrange bracket in \cite{RT6,contactBH}. It follows that the Lagrange bracket of a constant with a function may not in general be zero, i.e. $\{1,f\} = \xi(f)$ for any $f  \in C^\infty(\mathcal{M}_c,\mathbb{R})$, unlike the case with Poisson brackets wherein the bracket vanishes identically. This stems from the fact that $\{\cdot,\cdot\}$ does not satisfy the Leibniz condition and as a consequence, $\{\cdot,h\}$ does not define a vector field, i.e. it does not act as a derivation. It is easy to check that if the functions $f$ and $g$ are independent of $s$, the Lagrange bracket reduces to the Poisson bracket, as is anticipated. It can be straightforwardly shown that a contact vector field has the following form in terms of the Lagrange bracket:
\begin{equation}
  X_h(\cdot) = \{\cdot,h\} - (\cdot) \xi (h).
\end{equation} 
In this case $X_h (c) = 0$ where $c \in \mathbb{R}$ is some constant number.  

\section{Virial theorem for a contact Hamiltonian system}\label{VirSec}
In this section, we will derive a geometric version of the virial theorem suited for contact Hamiltonian systems following the earlier works \cite{virial1,virial2,virial3,virial4}. We can now state the following result. 

\begin{theorem}
Consider a contact Hamiltonian system $(\mathcal{M}_c,\eta,h)$ where $(\mathcal{M}_c,\eta)$ is a contact manifold and $h \in C^\infty(\mathcal{M}_c,\mathbb{R})$ is a function. Then, if $\phi_t$ be the flow generated by the corresponding Hamiltonian vector field $X_h$ where $t$ is an affine parameter called `time', then for an arbitrary observable $f \in C^\infty(\mathcal{M}_c,\mathbb{R})$ which remains bounded in its evolution along $X_h$, one has 
\begin{equation}\label{vir1}
  \langle \{f,h\} -f \xi (h) \rangle = 0,
\end{equation}
where $\{\cdot,\cdot\}$ is the Lagrange bracket on $(\mathcal{M}_c,\eta)$ and $\langle \cdot \rangle$ denotes time-averaging.
\end{theorem}

\textit{Proof -} If $\phi_t$ be the flow generated by $X_h$ where $t$ is an affine parameter, then clearly $\phi_t$ commutes with $X_h$. Consequently, one can write for any function $f \in C^\infty(\mathcal{M}_c,\mathbb{R})$,
\begin{equation}
   \frac{d}{dt} (\phi^*_t f) = X_h (\phi^*_t f) = \phi^*_t (X_h (f)) = \phi^*_t (\{f,h\} - f \xi (h)).
\end{equation}
Integrating this equation from $t = 0$ to $t = T$ and dividing both sides by $T$ gives
\begin{equation}
  \frac{1}{T} [f \circ \phi_T - f \circ \phi_0] = \frac{1}{T} \int_{0}^{T} dt (\{f,h\} - f \xi (h)) \circ \phi_t.
\end{equation}
We assume that upon taking the limit $T \rightarrow \infty$, the right-hand side does not blow up. This holds if the function $f$ remains bounded along its time evolution. Then we get Eq. (\ref{vir1}) upon taking $T \rightarrow \infty$. 

\vspace{2mm}

Eq. (\ref{vir1}) is a generalization of the result obtained in \cite{virial1} to the contact Hamiltonian framework. 

\begin{corollary}
If we choose $f= q^i p_i = G$, i.e. the virial of a mechanical system, then Eq. (\ref{vir1}) reduces to
\begin{equation}\label{Virialtheoremcontact}
  \langle \{G,h\}_{\rm PB} \rangle = \langle G \xi(h) \rangle,
\end{equation} where the subscript ${\rm PB}$ implies that $\{\cdot,\cdot\}_{\rm PB}$ is a Poisson bracket on the (exact) symplectic manifold $(\mathcal{M}_s,\omega = d\theta)$ which arises as the codimension one submanifold of $(\mathcal{M}_c,\eta)$, i.e. $\Phi: \mathcal{M}_s \mapsto \mathcal{M}_c$  such that $\theta = \Phi^* \eta$. 
\end{corollary}

\textit{Proof -} By direct calculation in local coordinates.

\vspace{2mm}

Notice that the Poisson bracket of two functions $f$ and $g$ reads
\begin{equation}
\{f,g\}_{\rm PB} = \frac{\partial f}{\partial q^i} \frac{\partial g}{\partial p_i} - \frac{\partial f}{\partial p_i} \frac{\partial g}{\partial q^i},
\end{equation} where $(q^i,p_i)$ are the Darboux coordinates on $(\mathcal{M}_s,\omega = d\theta)$, which is a codimension one submanifold of $(\mathcal{M}_c,\eta)$. Eq. (\ref{Virialtheoremcontact}) matches with the geometric version of virial theorem derived in \cite{virial1} for a symplectic manifold if $\xi(h) = 0$, i.e. if $h$ is independent of $s$. We discuss some examples below. 

\subsection{Damped oscillator}\label{dampedsub}
Let us consider the example of a one-dimensional damped harmonic oscillator with damping constant $\gamma$. The equation of motion is
\begin{equation}
\ddot q+\gamma\, \dot q+\omega^2\, q=0,\quad \quad  0<\omega,\gamma\in\mathbb{R}, \label{dhoeq}
\end{equation}
which is a particular example of a Li\'enard equation of first kind \cite{AL} satisfying the Chiellini condition \cite{Ch}. This problem was studied in detail in \cite{bateman,Cal41,Ka48}, both in the Lagrangian and Hamiltonian formalisms. In particular, the dynamics can be derived from a time-dependent Lagrangian (or Hamiltonian). The time-dependent Lagrangian
\begin{equation}
L(q,\dot{q},t)=\frac{1}{2}  e^{\gamma t}\left(m\dot{q}^2- m \omega^2\, q^2\right), \label{dhoLag}
\end{equation} leads to Eq. (\ref{dhoeq}) via the usual Euler-Lagrange equation \cite{goldsteincm}. This Lagrangian was proposed by Bateman \cite{bateman} and later by Caldirola \cite{Cal41}. Upon defining $p = \frac{\partial L}{\partial \dot{q}} = m e^{\gamma t} \dot{q}$, the Hamiltonian turns out to be
\begin{equation}
H(q, p, t) =\frac{p^2}{2m} e^{-\gamma t}+ \frac{m}{2} e^{\gamma t} \omega^2\, q^2, \label{dhoHam}
\end{equation} 
which corresponds to a time-dependent mass harmonic oscillator with mass $m(t)= m e^{\gamma t} $ \cite{Ka48}. Moreover, there are various other ways for describing Eq. (\ref{dhoeq}) such as via non-standard Lagrangians \cite{PGNS}, or conformal vector fields \cite{conformal}. In what follows, we shall be describing it using contact Hamiltonian dynamics. 

\vspace{2mm}

Its motion can be described by a contact Hamiltonian function of the form \cite{CM1,CM2,CM3}
\begin{equation}
  h = \frac{p^2}{2m} + \frac{m \omega^2 q^2}{2} + \gamma s ,
\end{equation} where $\gamma > 0$. The contact equations of motion read
\begin{equation}
  \dot{q} = \frac{p}{m}, \quad \dot{p} = - m \omega^2 q - \gamma p, \quad \dot{s} = \frac{p^2}{2m} - \frac{m \omega^2 q^2}{2} - \gamma s .
\end{equation} The first two equations give the equation of motion of a one-dimensional damped harmonic oscillator. In this case the virial is simply $G = q p$ and therefore, Eq. (\ref{Virialtheoremcontact}) gives
\begin{equation}\label{abc}
  \bigg\langle \frac{p^2}{2m} \bigg\rangle = \bigg\langle \frac{m \omega^2 q^2}{2} \bigg\rangle + \frac{1}{2} \langle \gamma p q \rangle.
\end{equation} 
This is exactly what one would have obtained using the Newton's second law: the term on the left-hand side is the mean kinetic energy, the first term on the right-hand side is the mean spring potential energy while the second term on the right-hand side is of the form $ -\frac{\langle \mathbf{r} \cdot \mathbf{F}\rangle}{2}$ [Eq. (\ref{1st})] where $F = -\gamma p$ is the non-potential force of linear friction. The above setting can be extended to any number of particles moving in an arbitrary potential $V(q^i)$ while experiencing a friction proportional to the momentum $p_i$ with damping constant $\gamma$. The result is
\begin{equation}
 \sum_i \bigg\langle \frac{p_i^2}{2m} \bigg\rangle = \frac{1}{2} \sum_i \bigg\langle q^i \frac{\partial V(q^i)}{\partial q^i} \bigg\rangle + \frac{1}{2} \sum_i \langle \gamma q^i p_i \rangle.
\end{equation}

\subsection{Parachute equation}\label{parachutesec}
Consider the case of vertical motion of a particle of mass $m$ falling through a fluid under constant gravity. The dynamics can be described by a contact Hamiltonian vector field. Following the Lagrangian description presented in \cite{CM3}, we consider the contact Hamiltonian function
\begin{equation}
h = \frac{(p - 2 \lambda s)^2}{2m} + \frac{mg}{2 \lambda}(e^{2 \lambda q} - 1),
\end{equation} where $q$ is the mechanical coordinate of the falling particle, $p$ is the generalized momentum, while $\lambda > 0$ is a constant. It can be shown that the frictional force is proportional to the square of the velocity. The contact equations of motion are
\begin{eqnarray}
\dot{q} &=& \frac{p - 2 \lambda s}{m}, \label{pa}\\
\dot{p} &=&- mg e^{2 \lambda q} + \frac{2 \lambda p(p - 2\lambda s)}{m}, \label{pb} \\
\dot{s} &=& \frac{p(p-2 \lambda s)}{m} - \frac{(p - 2 \lambda s)^2}{2m} - \frac{mg}{2 \lambda}(e^{2 \lambda q} - 1). \label{pc}
\end{eqnarray}
Eq. (\ref{pa}) gives
\begin{equation}
m \ddot{q} = \dot{p} - 2 \lambda \dot{s},
\end{equation} which using Eqs. (\ref{pb}) and (\ref{pc}) leads to
\begin{equation}\label{eompara}
 \ddot{q} = \lambda  \dot{q}^2  - g ,
\end{equation} matching with the second-order equation of motion obtained in \cite{CM3} via the contact Lagrangian method. Notice that the drag has a quadratic dependence on velocity unlike the system discussed in the previous subsection where friction was linear in velocity. The constant $\lambda$ depends on the density of the fluid, the shape of the object, etc. Strictly, the hypothesis that $G$ remains bounded is not satisfied for unbounded fall and the time-average relation should therefore be interpreted either on finite time windows, or for bounded motions.

\vspace{2mm}

Now, for the present case of a particle falling under gravity while experiencing a quadratic drag, Eq. (\ref{Virialtheoremcontact}) gives
\begin{equation}\label{virthpara}
\bigg\langle \frac{p(p - 2 \lambda s)}{m} \bigg\rangle = - 2 \lambda \bigg\langle \frac{pq ( p - 2 \lambda s)}{m} \bigg\rangle + mg \langle q e^{2 \lambda q} \rangle,
\end{equation}
where one can immediately verify that for $\lambda = 0$ (no dissipation), one has $\langle p^2/m \rangle = \langle mg q \rangle$, which is the result one would expect from the virial theorem for a particle falling under constant gravity. One notices that for $\lambda \neq 0$, the virial theorem [Eq. (\ref{virthpara})] doesn't appear as if it presents a relationship between the time-averages of energies. This is because in this case $p$ is not the mechanical momentum of the particle. On the other hand, if one were to use a Lagrangian framework as will be done in Sec. (\ref{parlagsec}), then the corresponding virial theorem would indeed relate the time-averaged kinetic and potential energies. 

\subsection{Forced oscillator}\label{Forsec}
In Sec. (\ref{dampedsub}), we considered the case of a particle acted upon by a damping force proportional to its velocity. We will now consider the case of a forced oscillator where a periodic forcing continues to drive the motion as $T \rightarrow \infty$. For this, we require a contact Hamiltonian function of the following form 
\begin{equation}\label{tdf}
  h = \frac{p^2}{2m} + \frac{m \omega^2 q^2}{2} + \gamma s - q \mathcal{F}(t),
\end{equation} where $\gamma > 0$ and $\mathcal{F}(t)$ is the external periodic force.

\vspace{2mm}

Let us notice that the contact Hamiltonian is now explicitly time-dependent. Therefore, we use the machinery discussed in Sec. (\ref{TDsec}) to describe the dynamics. Following the previous treatments, and with the identification $G = qp$, one can reproduce after time-averaging, Eq. (\ref{Virialtheoremcontact}) describing the generalized virial theorem. 

\vspace{2mm}
 
Let us consider the case where $\mathcal{F}(t) = F_0 \cos (\Omega t)$, where $F_0$ and $\Omega$ are positive constants. The equations of motion are 
\begin{eqnarray}
 \dot{t} &=& 1, \\
 \dot{q} &=& \frac{p}{m}, \\
 \dot{p} &=& - \gamma p - m \omega^2 q  +  F_0  \cos (\Omega t) ,\\
 \dot{s} &=& \frac{p^2}{2m} - \frac{m \omega^2 q^2}{2} - \gamma s + q F_0  \cos (\Omega t).
\end{eqnarray}
The second and third equations above describe the damped and forced oscillator. Subsequently, Eq. (\ref{Virialtheoremcontact}) gives
\begin{equation}
\bigg\langle \frac{p^2}{2m} \bigg\rangle = \bigg\langle \frac{m \omega^2 q^2}{2} \bigg\rangle - \frac{1}{2}\big\langle q \big[-\gamma p + F_0 \cos (\Omega t)\big] \big\rangle,
\end{equation} where the right-hand side contains two terms: the first one is the usual potential energy contribution from the `spring', while the second one is due to the non-potential force and is of the type $ -\frac{\langle \mathbf{r} \cdot \mathbf{F}\rangle}{2}$ as in Eq. (\ref{1st}). 
 
\subsection{Brownian oscillator}\label{Bsec}
We consider the case of Brownian motion wherein, a single particle of mass $m$ is immersed in a fluid where it experiences both random collisions from the fluid particles and a viscous drag. Furthermore, there is a harmonic trap in which the particle moves. The random force $f(t)$ is termed as `noise' which has the following statistical properties:
\begin{equation}\label{noisestat}
\langle f(t) \rangle = 0, \quad \langle f(t) f(t') \rangle = 2 m \gamma k_B \mathcal{T} \delta(t-t'),
\end{equation} where $\mathcal{T}$ is the absolute temperature and the averaging above is a thermal-average that is consistent with time-averaging due to ergodic hypothesis. In order to describe it in the framework of time-dependent contact Hamiltonian dynamics [Sec. (\ref{TDsec})], we consider a contact Hamiltonian function of the form given in Eq. (\ref{tdf}) with $\mathcal{F}(t) = f(t)$. The resulting equations of motion are 
 \begin{eqnarray}
 \dot{t} &=& 1, \\
 \dot{q} &=& \frac{p}{m}, \label{lang1} \\
 \dot{p} &=& - \gamma p - m \omega^2 q  +  f(t), \label{lang2}  \\
 \dot{s} &=& \frac{p^2}{2m} - \frac{m \omega^2 q^2}{2} - \gamma s + q f(t).
 \end{eqnarray}
 Eqs. (\ref{lang1}) and (\ref{lang2}) can be combined to give the famous Langevin equation
 \begin{equation}\label{langevin}
  \ddot{q} + \gamma \dot{q} + \omega^2 q = f(t)/m.
 \end{equation}
The Brownian particle follows a zig-zag trajectory (in one dimension, the particle can go either left or right, but the `zig-zag' nature remains unchanged). In this situation Eq. (\ref{Virialtheoremcontact}) gives
 \begin{equation}
\bigg\langle \frac{p^2}{2m} \bigg\rangle = \bigg\langle \frac{m \omega^2 q^2}{2} \bigg\rangle -\frac{1}{2}\big\langle q \big[-\gamma p + f(t)\big] \big\rangle.
 \end{equation}
 If one solves the Langevin equation \cite{z,b}, then using the solution $(q(t),p(t))$, it can be shown that $\langle q(-\gamma p + f(t)) \rangle = 0$, where one uses Eq. (\ref{noisestat}) describing the statistical properties of the noise. This gives the result that the time-averaged kinetic and potential energies of the Brownian oscillator are equal. Moreover, using the solution $(q(t),p(t))$ and Eq. (\ref{noisestat}), one can show that they are equal to $k_B \mathcal{T}/2$, in accordance with the classical equipartition theorem. 

\subsection{Gierer-Meinhardt system}
The dynamics of the autocatalytic Gierer-Meinhardt system \cite{GM} can be obtained via conformal as well as contact Hamiltonian dynamics \cite{virial4,PG}. To recall, on a symplectic manifold $(\mathcal{M}_s,\omega)$, a conformal vector field $X_H^a$ associated with some Hamiltonian $H \in C^\infty(\mathcal{M}_s,\mathbb{R})$ for some constant $a \in \mathbb{R}$ is one which gives $\pounds_{X_H^a} \omega = a \omega$. For $a \neq 0$, this corresponds to canonical, but not strictly canonical, transformations \cite{conformal}. The particular case $a = 0$ gives a genuine Hamiltonian vector field $X_H$ which is volume preserving. We shall consider exact symplectic manifolds, i.e. for which $\omega$ is exact, or equivalently, one has $\omega = d\theta$. Then, it is easy to verify that the vector field $X_H^a$ satisfies 
\begin{equation}
 \iota_{X_H^a} \omega = dH + a \theta.
\end{equation} 
Let us define a vector field $Z$ which satisfies $\iota_Z \omega = \theta$. Since $\pounds_{X_H} \omega = 0$ and $\pounds_Z \omega = \omega$, one has $\pounds_{X_H + aZ} \omega = a \omega$ thereby indicating that one can write $X_H^a = X_H + a Z$ \cite{conformal}. 

\vspace{2mm}
 
The Gierer-Meinhardt system is associated with the dynamical equations
\begin{equation}\label{GME}
 \dot{x} = \frac{A}{B + y} - Cx, \quad \quad \dot{y} = Dx - Ky,
\end{equation} where $A, B, C, D, K$ are constants. If one considers a two-dimensional symplectic phase space $\mathcal{M}_s = T^* \mathbb{R} \simeq \mathbb{R}^2$ with $\omega = dx \wedge dy$ and $\theta = -y dx$, then Eq. (\ref{GME}) can arise as the equations whose solutions are the integral curves of the conformal vector field $X^{-(C + K)}_H$ and the associated Hamiltonian function reads
\begin{equation}
 H = A \ln (B + y) - \frac{Dx^2}{2} - Cxy.
\end{equation} More explicitly, the equations of motion are obtained from $H$ given above as
\begin{equation}
 \dot{x} = \frac{\partial H}{\partial y}, \quad \quad \dot{y} = - \frac{\partial H}{\partial x} - (C + K) y. 
\end{equation}
The dynamics of the Gierer-Meinhardt system can be equivalently described using contact Hamiltonian dynamics (see also \cite{PG}) if we consider an extended phase space $\mathcal{M}_c = T^*\mathbb{R} \times \mathbb{R} \simeq \mathbb{R}^3$ with a contact form $\eta = dz - y dx$. The appropriate choice of contact Hamiltonian is 
\begin{equation}
 h = A \ln (B + y) - \frac{Dx^2}{2} + C(z - xy) + K z,
\end{equation} which implies that the contact equations give Eq. (\ref{GME}) and additionally, an equation of motion for $z$:
\begin{equation}
 \dot{z} = A \Bigg[ \frac{y}{B+y} - \ln (B + y) \Bigg] + \frac{Dx^2}{2} - (C+K)z.
\end{equation}  
Geometrically, one may say that the Gierer-Meinhardt dynamics on the two-dimensional phase space $\mathcal{M}_s = T^*\mathbb{R}$ is a projection of $X_h$ defined on the three-dimensional space $\mathcal{M}_c = T^*\mathbb{R} \times \mathbb{R}$. This is true in a more general setting: Consider an exact symplectic manifold $(\mathcal{M}_s,\omega)$ as a codimension one submanifold of a contact manifold $(\mathcal{M}_c,\eta)$ such that locally, $\mathcal{M}_c = \mathcal{M}_s \times \mathbb{R}$ wherein, for the inclusion map $\Phi: \mathcal{M}_s \mapsto \mathcal{M}_c$, one has $\theta = \Phi^* \eta$, where $\theta$ is the symplectic potential on $\mathcal{M}_s$. Then, any conformal vector field $X_H^a$ in $(\mathcal{M}_s,\omega)$ with $H \in C^\infty(\mathcal{M}_s,\mathbb{R})$ can be obtained as a projection of a contact vector field $X_h$ on $(\mathcal{M}_c, \eta)$ for $h \in C^\infty(\mathcal{M}_c,\mathbb{R})$ where, $h = H - as$ with $s \in \mathbb{R}$. 

\vspace{2mm}
 
Since the system being considered at present is not a mechanical system, there is no obvious identification of the virial $G$. Nevertheless, putting $G = xy$ in analogy with a mechanical system gives rise to the equality
\begin{equation}\label{GMVir}
  \bigg\langle \bigg(1 + \frac{B}{y}\bigg)^{-1} \bigg\rangle + \alpha \langle x^2 \rangle = \beta \langle xy \rangle,
\end{equation}
where $\alpha = \frac{D}{A}$ and $\beta = \frac{C+K}{A}$. Eq. (\ref{GMVir}) can be termed as the virial theorem for the Gierer-Meinhardt system. 
 
 \section{Contact Lagrangian framework}\label{LagSec}
We will, for the sake of completeness, derive a generalized virial theorem in the contact Lagrangian framework \cite{CM3,CM4}. Consider the bundle $\pi: TQ \times \mathbb{R} \rightarrow Q$ and a Lagrangian function $L: TQ \times \mathbb{R} \rightarrow \mathbb{R}$. On $TQ$, one has the usual endomorphism $S$ which can be naturally extended to $TQ \times \mathbb{R}$ and in a local coordinate chart $(q^i,\dot{q}^i,s)$ one has
\begin{equation}
 S = dq^i \otimes \frac{\partial}{\partial \dot{q}^i}.
\end{equation}
The local coordinates are such that $(q^i,\dot{q}^i)$ are the coordinates on $TQ$: $q^i$ are the coordinates on $Q$ and $\dot{q}^i$ are the fiber coordinates, whereas, $s$ is a global coordinate in $\mathbb{R}$. One may now construct a one-form $\theta_L$ as $\theta_L = S^*(dL)$ so that
\begin{equation}\label{etaL}
\eta_L = ds - \theta_L = ds- \frac{\partial L}{\partial \dot{q}^i} dq^i,
\end{equation} defines a contact form, meaning that $\eta_L \wedge (d\eta_L)^n \neq 0$, provided $L$ is regular, i.e. the matrix $W$ with elements $W_{ij}= \frac{\partial^2 L}{\partial \dot{q}^i \partial \dot{q}^j}$ is invertible. The Reeb vector field $\xi_L$ is defined via the conditions $\eta_L(\xi_L)=1$ and $\iota_{\xi_L}d\eta_L = 0$, taking the following form in local coordinates:
\begin{equation}
  \xi_L = \frac{\partial}{\partial s} - W^{ij}  \frac{\partial^2 L}{\partial \dot{q}^j \partial s} \frac{\partial}{\partial \dot{q}^i},
\end{equation} where $W^{ij}$ are the matrix elements of the inverse of $W$, i.e. $W^{-1}$ which exists due to the regularity of $L$. Now, the energy function $E_L: TQ \times \mathbb{R} \rightarrow \mathbb{R}$ is defined as $ E_L = \Delta(L) - L$, where $\Delta$ is the natural extension to $TQ \times \mathbb{R}$ of the Liouville vector field $\Delta = \dot{q}^i \frac{\partial}{\partial \dot{q}^i}$ on $TQ$. This means 
\begin{equation}\label{EL}
 E_L = \dot{q}^i \frac{\partial L}{\partial \dot{q}^i} - L. 
\end{equation} We have mentioned all the ingredients required to describe contact Lagrangian dynamics on $TQ \times \mathbb{R}$. Corresponding to $L$, we define a vector field $X_L$ which satisfies
\begin{equation}\label{XLCond}
 \eta_L (X_L) = - E_L, \quad \quad \iota_{X_L} d\eta_L = dE_L - (\pounds_{\xi_L} E_L) \eta_L,
\end{equation} and is therefore, a contact Hamiltonian vector field associated with the energy function $E_L$. One has the following local expression for $X_L$: 
\begin{equation}\label{XLExp}
 X_L = L \frac{\partial}{\partial s} + \dot{q}^i \frac{\partial}{\partial q^i} + W^{ik} \bigg( \frac{\partial L}{\partial q^k} - \frac{\partial^2 L}{\partial q^j \partial \dot{q}^k} \dot{q}^j - L \frac{\partial^2 L}{\partial s \partial \dot{q}^k} + \frac{\partial L}{\partial s} \frac{\partial L}{\partial \dot{q}^k} \bigg) \frac{\partial}{\partial \dot{q}^i}.
\end{equation}
It follows that the integral curves of the vector field $X_L$ satisfy the generalized Euler-Lagrange equations obtained by Herglotz in 1930 via a variational principle \cite{Herglotz} (see also \cite{CM4}). 

\vspace{2mm}

We shall now discuss the generalized virial theorem in the Lagrangian setting. To summarize the above discussion, given any configuration space $Q$, a regular Lagrangian $L: TQ\times \mathbb{R} \rightarrow \mathbb{R}$ induces a contact structure on $TQ \times \mathbb{R}$ with a contact form $\eta_L$ defined in Eq. (\ref{etaL}). The energy function is defined as in Eq. (\ref{EL}) and corresponding to the Lagrangian $L$ one may define a dynamical vector field $X_L$ via the conditions given in Eq. (\ref{XLCond}) whose local expression is Eq. (\ref{XLExp}). The solutions of this contact Lagrangian problem solve the generalized Euler-Lagrange equations of Herglotz.

\vspace{2mm}

\begin{theorem}
Consider a contact Lagrangian system with some $L:TQ	\times \mathbb{R} \rightarrow \mathbb{R}$ such that $X_L$ is the dynamical vector field. Then, the following result holds:
\begin{equation}\label{virialLag}
 \langle X_L (G) \rangle = 0,
\end{equation} where $G = m_{ij} \dot{q}^i q^j$ is the virial and $\langle \cdot \rangle$ denotes time-average. 
\end{theorem}

\textit{Proof -} For a contact Lagrangian system, the dynamical vector field is $X_L$ which means the dynamics of any function $f:TQ \times \mathbb{R} \rightarrow \mathbb{R}$ is given by $X_L (f)$. If $\phi^L_t$ be the flow generated by $X_L$ where $t$ is time, then $\phi^L_t$ commutes with $X_L$ and therefore 
\begin{equation}
   \frac{d}{dt} (\phi^{L*}_t f) = X_L (\phi^{L*}_t f) = \phi^{L*}_t (X_L (f)).
\end{equation}
Taking time-average with $T \rightarrow \infty$ therefore gives $ \langle X_L (f) \rangle = 0$ where we have assumed that $f$ remains bounded in its time evolution. Thereafter, picking $f = G$ (the virial) gives Eq. (\ref{virialLag}) which can be interpreted as a generalized virial theorem in the contact Lagrangian framework. For a system with several generalized coordinates, one picks $G = m_{ij} \dot{q}^i q^j$ where $m_{ij}$ is a suitable mass matrix which is symmetric by definition\footnote{The mass matrix generalizes the notion of mass. For pure translational motion, it is typically a diagonal matrix with the masses of the particles appearing as diagonal entries. For rotational motion, it corresponds to the moment of inertia tensor, and therefore its components may depend upon $\{q^i\}$. The mass matrix is symmetric because the kinetic energy is defined as $K = \frac{1}{2} m_{ij} \dot{q}^i \dot{q}^j$.}. In the simple case of a one-particle system undergoing translational motion, $m_{ij} = m \delta_{ij}$, where $m$ is the mass of the particle and $\delta_{ij}$ is the Kronecker delta.  

\vspace{2mm}

It should be remarked that the result $ \langle X_L (f) \rangle = 0$ is true for any function $f$ that remains bounded in its time evolution but the choice $f = G$ gives rise to an appropriate virial theorem. Below, we consider two examples. 

\subsection{Damped oscillator}
Consider the one-dimensional damped oscillator whose Hamiltonian formalism was discussed in Sec. (\ref{dampedsub}). The associated contact Lagrangian is of the form
\begin{equation}
L = \frac{m \dot{q}^2}{2} - \frac{m \omega^2 q^2}{2} - \gamma s.
\end{equation} 
Then, the dynamical vector field [Eq. (\ref{XLExp})] becomes
\begin{equation}
X_L = L \frac{\partial}{\partial s} + \dot{q} \frac{\partial}{\partial q} - \big( \omega^2 q  + \gamma \dot{q}\big) \frac{\partial}{\partial \dot{q}},
\end{equation} which means the equations of motion are $\dot{s} = L$ and, 
\begin{equation}
\ddot{q} = - \big( \omega^2 q  + \gamma \dot{q}\big).
\end{equation}
If we identify $G = m \dot{q} q $, then the generalized virial theorem [Eq. (\ref{virialLag})] gives
\begin{equation}
\bigg\langle \frac{m\dot{q}^2}{2} \bigg\rangle = \bigg\langle \frac{m \omega^2 q^2}{2} \bigg\rangle + \frac{1}{2}\langle m \gamma q \dot{q} \rangle.
\end{equation}
Thus, the term on the left-hand side gives the time-averaged kinetic energy while the first term on the right-hand side gives the time-averaged spring potential energy. The second term on the right-hand side is of the form $-\frac{\langle \mathbf{r} \cdot \mathbf{F}\rangle}{2}$ appearing in Eq. (\ref{1st}) where $F$ is the non-potential frictional force $F = - m \gamma \dot{q}$. 

\subsection{Parachute equation}\label{parlagsec}
We will consider another example, namely the parachute equation whose Hamiltonian formalism was discussed in Sec. (\ref{parachutesec}). The following Lagrangian describes the dynamics \cite{CM3}:
\begin{equation}
L = \frac{m \dot{q}^2}{2} - \frac{mg}{2 \lambda} \big( e^{2 \lambda q} - 1 \big) + 2 \lambda \dot{q} s,
\end{equation} for which simple manipulations show that the dynamical vector field is
\begin{equation}
X_L = L \frac{\partial}{\partial s} + \dot{q} \frac{\partial}{\partial q} + (\lambda \dot{q}^2 - g) \frac{\partial}{\partial \dot{q}}.
\end{equation}
The equations of motion give rise to Eq. (\ref{eompara}) describing the acceleration of a particle of mass $m$ which falls under gravity through a fluid leading to a drag force quadratic in velocity. For this case, identifying $G = m \dot{q} q $, the generalized virial theorem [Eq. (\ref{virialLag})] gives 
\begin{equation}
\bigg\langle \frac{m\dot{q}^2}{2} \bigg\rangle = \frac{1}{2}\langle m g q \rangle - \frac{1}{2}\langle m \lambda q \dot{q}^2 \rangle,
\end{equation} wherein the first term on the right-hand side describes the time-averaged gravitational potential energy (with an overall 1/2 factor) while the second term consistently gives $-\frac{\langle \mathbf{r}\cdot \mathbf{F} \rangle}{2}$ for the non-potential quadratic drag force $F = m \lambda \dot{q}^2$. Note that the hypothesis that $G$ remains bounded is not satisfied for unbounded fall and the time-average relation should be interpreted either on finite time windows, or for bounded motions.

\section{Conclusions} \label{DisSec}
In this paper, we studied the generalized virial theorem for contact Hamiltonian dynamics and discussed various examples including mechanical systems with dissipation. Although the result comes out to be distinct from that obtained earlier specific to symplectic Hamiltonian vector fields, it is found to give consistent results for mechanical systems. It also follows that if the contact Hamiltonian is independent of $s$, i.e. $\xi(h) = 0$, then the generalized virial theorem obtained here reduces to that for a symplectic Hamiltonian system \cite{virial1,virial2,virial3,virial4}, where the phase space is a codimension one symplectic submanifold of the contact manifold we started with. Subsequently, we studied the generalized virial theorem for a contact Lagrangian system where a regular Lagrangian $L:TQ \times \mathbb{R} \rightarrow \mathbb{R}$ leads to a contact form on $TQ \times \mathbb{R}$. Two mechanical examples, namely particle motion under linear dissipation and motion under quadratic drag were discussed. 

\vspace{2mm}

There are several avenues for future developments. It would be nice to develop an understanding of the generalized virial theorem in the context of thermodynamics, given that contact Hamiltonian systems describe thermodynamic transformations, both reversible \cite{RT5} (see also \cite{contactBH} and references therein) and irreversible \cite{generic1,generic2,generic4}. Moreover, it would be interesting to study a quantum version of the generalized virial theorem specific to contact manifolds where dissipation is inbuilt. For instance, in \cite{CM2} the possibility of a `contact Schr\"odinger equation' was suggested which could describe certain dissipative quantum systems such as the Schr\"odinger-Langevin equation \cite{SL,SL1}. It would therefore be very interesting to generalize the results of this paper to quantum mechanical situations, particularly to those involving open systems. We hope such issues will be reported on, in the future. 

\section*{Acknowledgements}
The author is supported by the Ministry of Education (MoE), Government of India in the form of a Prime Minister's Research Fellowship (ID: 1200454).

\end{document}